# Mapping of Axial Strain in InAs/InSb Heterostructured Nanowires

Atanu Patra, Jaya Kumar Panda, and Anushree Roy[*],
[1]*Department of Physics, Indian Institute of Technology Kharagpur, Kharagpur 721 302, India*
Mauro Gemmi and Jérémy David,
[2]*Center for Nanotechnology Innovation @ NEST, Istituto Italiano di Tecnologia, Piazza S. Silvestro 12, I-56127 Pisa, Italy*
Daniele Ercolani and Lucia Sorba[†],
[3]*NEST-Istituto Nanoscienze-CNR and Scuola Normale Superiore, Piazza S. Silvestro 12, I-56127 Pisa, Italy*

The article presents a mapping of the residual strain along the axis of InAs/InSb heterostructured nanowires. Using confocal Raman measurements, we observe a gradual shift in the TO phonon mode along the axis of these nanowires. We attribute the observed TO phonon shift to a residual strain arising from the InAs/InSb lattice mismatch. We find that the strain is maximum at the interface and then monotonically relaxes towards the tip of the nanowires. We also analyze the crystal structure of the InSb segment through selected area electron diffraction measurements and electron diffraction tomography on individual nanowires.

---

[*] Email: anushree@phy.iitkgp.ernet.in
[†] Email: lucia.sorba@sns.it

In recent times semiconductor research and development sectors embraced the use of III-V semiconductor nanowires (NWs) for high speed devices,[1] high frequency electronics,[2] and spin related applications.[3] Among these NWs, InSb NWs draw a special attention. From the basic physics point of view, finding Majorana fermions in InSb NWs in the presence of s-wave superconductors[4] or the manipulation of strong spin-orbit coupled qubit states,[5] have led to new dimensions in the research on this system.

The main hindrance in using lattice mismatched semiconductor nanostructures in any application is the occurrence of strain-induced defects in the system. However, it is claimed that the interfacial strain, which mostly appears in lattice mismatch of the NW materials, relaxes within few nanometers across the interface.[6,7,8,9] For example, in a recent report,[6] using scanning transmission electron microscopy along with computer simulation and geometric phase analysis, it has been demonstrated that in InAs/InSb NWs the interfacial strain relaxes within 10 nm via both elastic deformation (misfit dislocations) and plastic deformation (plane bending). The misfit dislocations are observed at the center of the NWs, whereas the interfacial strain relaxes via the plane bending near the edge of the NWs.

In above references, electron diffraction measurements have been used to quantify the strain in NWs. Though, it is a direct probe to estimate strain by measuring lattice distortion, the resolution of the micrographs limits the measurable strain in a crystal structure. In general, using this technique, one is able to measure the strain of more than 1% in the crystal structure of semiconductor NWs.[8,10] In comparison to electron diffraction measurements, Raman spectroscopy is more sensitive to the local strain in the crystal structure. However, the far field diffraction of light limits the spatial resolution of Raman scattering measurements while mapping the strain along the axis of the NWs. In this article we employ confocal Raman spectroscopy to measure the strain in the crystal structure, along the axis of the InAs/InSb NWs, grown by chemical beam epitaxy technique. The selected area electron diffraction (SAED) measurements and electron diffraction tomography (EDT) are exploited to measure the extent of lattice distortion, on individual NW.

InAs/InSb NW growths were performed by Au-assisted CBE in a Riber Compact 21 system. The system employs pressure control in the metalorganic (MO) lines to determine precursor fluxes during the sample growth. The precursors involved in the NW growth are tri-methylindium (TMIn), tertiarybutylarsine (TBAs) and trisdimethylaminoantimony (TDMASb). A nominally 1 nm–thick Au film was first deposited on (111)B InAs wafers by thermal evaporation. Before the growth was initiated, the samples were heated at 560±10 °C under TBAs flow for 20 min in order to dewet the Au film into nanoparticles and to remove the surface oxide from the InAs substrate. InAs stems were grown at a temperature of 430±10 °C for 90 min with MO line pressures of 0.3 and 1 for TMIn and TBAs, respectively. The temperature was then lowered to 410±10 °C and InSb segments were grown for 120 min with MO line pressures of 0.4 and 0.7 for TMI and TDMASb, respectively. In order to reduce the lateral growth, the temperature was increased to 445±10 °C and the InSb segments were grown for another 270 min. Figure 1 shows a 45° tilted scanning electron micrograph of vertically aligned InAs/InSb NWs. The investigated InSb segments have zincblende crystal structure, a length of ~4.0 μm and a wide diameter distribution in the 100-200 nm range.

Confocal Raman measurements were carried out in back-scattering geometry. In a confocal microscope, a pin-hole is used in the optically conjugate plane in front of the detector to eliminate the out-of-focus light from the sample. Thus, it helps in improving the spatial resolution of the depth profiling by Raman scattering measurements. A pin-hole with a diameter of 50 μm has been used. We worked with a 100× objective lens (numerical aperture 0.9) to focus the laser beam on standing NWs. All Raman scattering results are the ensemble average for 4-5 NWs. A 488 nm line of Ar$^+$ laser was used as an excitation source. Laser power was kept at ~ 50 μW to avoid the effect of local heating. Raman spectra were recorded with a confocal microscope (model BX41, Olympus, Japan) and a triple monochromator (model T64000, JY, France).

Because of far field diffraction limit of light, even in confocal Raman spectroscopy it is necessary to calibrate the depth resolution of the measurements. We measure the ratio of intensity of Raman scattered light through the pinhole ($I_{CF}$) and without the pinhole ($I_0$), at different distances z from the focal plane of the objective lens. The FWHM of the plot of this ratio *vs* z provides the measure of the depth resolution for the given pinhole.[11] Figure 2 presents the measured step response curve for our experimental set-up, using a thick bulk sample. From this calibration we obtain a depth resolution of 1.2 μm.

Keeping the same experimental conditions, employed for the calibration, we investigated vertically aligned InAs/InSb NWs (shown in Figure 1). The laser light was first focused at the bottom of the InSb segment of the standing NWs (*i.e.* in the portion of the InSb segment near the interface with the InAs stem). The sample stage was then moved down by steps of 0.25 μm to access different points along the axis of the InSb segments towards the tip. Refer to the schematic diagram shown in the inset of Figure 2.

We performed the selected area electron diffraction (SAED) measurements on a Zeiss Libra 120 operating at 120 kV. The measurements have been performed with the Digital Micrograph software by Gatan, using intensity profiles along the [111] and [1-11] directions, both having equivalent cubic symmetry.

The evolution of Raman spectra along the length of the NWs is shown in Figure 3(a). The bottom-most spectrum was recorded when the light was focused on the part of InSb segment near the interface with the InAs stem, while the top-most spectrum was acquired near the tip of the NW (refer to the side panel of Figure 3(a)). At room temperature, TO and longitudinal optical (LO) phonon modes of bulk InSb appear at 179.7 and 191.0 cm-1.[12] Each spectrum in Fig. 3(a) has been deconvoluted with two Lorentzian functions for TO and LO modes, keeping peak positions, widths and intensities as free fitting parameters. Deconvoluted spectra are shown as magenta lines for the lowest spectrum of Figure 3(a). The vertical red dashed line in .figure 3(a), marks the position of the transverse optical (TO) peak of each curve to highlight its gradual shift in the spectra towards the tip of the NW. The Raman shift of the TO phonon mode along the length of the InSb NW, as obtained from this analysis, is shown in Figure 3(b). In the figure, we consider the bottom of the InSb segment near the interface with the InAs stem as the zero of the z-axis in Figure 3(b). Keeping in mind the resolution associated with our z-scan (shown in Figure 2), the horizontal error bar for each data point plotted in Fig. 3(b) has been kept as ±1.2 μm We observe a monotonous change in Raman shift of the TO mode as the laser probe moves from the InAs/InSb interface towards the tip of the InSb segment (positive z as per our notation). Similar shift in the TO phonon mode has been observed in the spectra acquired in z scans along the opposite direction, i.e. when the laser was first focused at the tip of the NWs and the sample stage was moved up to accesses the other points along the axis towards the InAs stem.

Several possible explanations can be argued in order to interpret the observed TO phonon mode shifts, First, as the InSb segments are grown on InAs stems, between InAs and InSb segments a transition region of $InAs_xSb_{1-x}$ may exist.[6] Thus, the observed shift in the phonon mode may be due to a small variable residual As concentration along the length of the NW. However, the formation of such alloy is, generally, confined within few nm near the heterointerface, while the length of our InSb segments is ~4 μm. Moreover, InSb-like TO peak position in $InAs_xSb_{1-x}$ does not change with As content x.[13] Thus, we rule out the possibility that the observed shift in the phonon mode in Figure 3(b) is due to a change in composition along the axis of the NW.

Second, the observed TO phonon shifts can be associated to phonon confinement effects. However, the diameter of the NWs, used in our study, is too large (~100-200 nm) to shift their Raman peak. Hence, even a slight variation in the diameter of the NWs, from the base to tip, would not shift the phonon peak. .

Another explanation could be associated to the presence of residual strain along the NWs. In fact, the phonon frequency of a crystal depends on the interatomic force constants. Any local distortion in a perfect crystal structure modifies the deformation potential-optical phonon interactions in semiconductors.[14] This short range interaction is manifested in relative displacement of atoms at the zone center optical phonons.[15] The strain tensor components, arising from the static distortion in the crystal, result in non-identical movement of the atoms, and modify the phonon frequencies. The modified TO frequency can be related to the hydrostatic stress and shear stress along [111] ZB plane by[15]

$$\Omega = \Omega_0 + \Delta\Omega_H + \frac{2}{3}\Delta\Omega_s, \text{ with}$$

$$\Delta\Omega_H = \frac{\sigma}{6\Omega_0}(p+2q)(S_{11}+2S_{12}) \text{ and } \Delta\Omega_s = \frac{\sigma}{2\Omega_0}rS_{44} \quad \ldots\ldots\ldots[1]$$

Here, $\Omega_0$ is the phonon frequency of the unstrained crystal. $\sigma$ is the net stress in the crystal. *p,q* and *r* coordinates are related to deformation potential and $S_{ij}$ are elastic compliance constants with respect to the cubic axis. The corresponding hydrostatic strain and shear strain in the crystal are

$$\varepsilon_H = (S_{11} + 2S_{12})\frac{\sigma}{3} \quad \text{and} \quad \varepsilon_S = \frac{S_{44}}{2}\frac{\sigma}{3} \quad \ldots\ldots\ldots\ldots\ldots(2)$$

$\varepsilon_H$ and $\varepsilon_S$ correspond to diagonal and off-diagonal components of the strain tensor matrix. Using Equations 1 and 2, along with the reported values of the physical constants ($S_{ij}$, *p, q, r*) for InSb,[13,16,17] and measured $\Delta\Omega = \Omega - \Omega_0$ from Figure 3(b), the variation of hydrostatic and shear strain along the axis of InSb NW are plotted in Figure 4. We find that both hydrostatic and shear strain do exist along the wire. They are maximum near the interface and gradually decrease along the axis of the NW. The measured strain varies from 0.06% to 0.28%.

Next we collected selected area electron diffraction (SAED) patterns at the bottom, center and top of the InSb segment in several NWs. Since electron diffraction has a poor resolution in determining the absolute value of the unit cell parameters (which could be related to the above measured weak strain), we focused the analysis on the determination of the distortion of the cubic crystal structure, which relays on relative measurements on equivalent parameters.

At first, we evaluated deformation of the structure in [011] zone axis pattern by comparing the *d* spacing along two equivalent (111) directions, one parallel and the other inclined to the NW axis (Figure 5(a)). We derived the two equivalent $d_{111}$ spacing by measuring the distances between Friedel pair reflections of the third order (*i.e.* 333 and -3-3-3, see arrows Figure 5(a)). In all cases the differences between the two *d*-spacing were within the experimental error while the maximum noticeable difference being 3.68±0.03 Å against 3.72±0.03 Å.

Next we measured the cell parameters on a 3-dimensional reconstruction of the reciprocal space (Figure 5(b)) obtained with the electron diffraction tomography technique (EDT).[18] In EDT a set of diffraction patterns are collected while the sample is rotated around the goniometer axis in steps of the order of 0.5° for a total angular range of 60°. It is possible to obtain the unit cell parameters from the position of the diffracted spots in the three dimensional reconstruction, derived from the recorded patterns with the PETS software[19]. To check possible distortions of the cubic structure we indexed the cubic face centered cell with a primitive rhombohedral cell, which should have—in case of an undistorted lattice— a=b=c and α=β=γ=60°. In our NWs the distortion in the lattice is found to be beyond the sensitivity of the technique. The maximum measured difference between the two lengths is 4.40±0.04 Å against 4.44±0.04 Å and between the two angles is 59.3±0.5° against 59.9±0.5°.

In summary, confocal Raman measurements, which is highly sensitive to the atomic-scale local disorder in a crystal structure, allowed us to observe a weak residual strain of ~0.2% in our NWs. The observed strain gradually decays from the interface towards their tip of the InAs/InSb NW. Conversely, the measured strain is extremely weak to be detected by SAED or EDT measurements, within the resolution limit of these experimental techniques. In fact, as most of the reports in the literature, claiming the strain relaxation at the interface in NWs, are based on electron diffraction techniques, the residual strain may not have been detected earlier. We believe that our studies may pave way for better understanding of strain release mechanisms in heterestructured NWs.

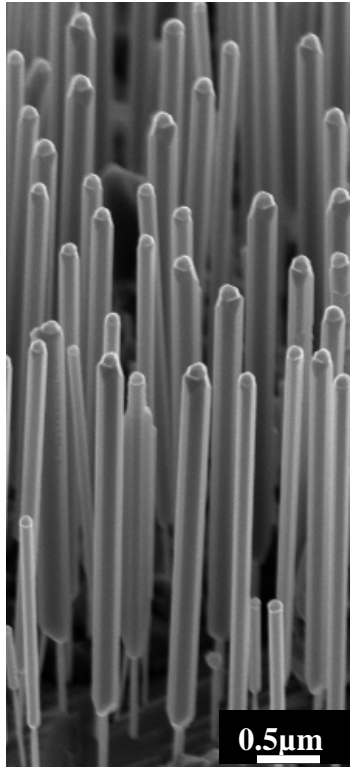

FIG. 1. 45° tilted SEM image of aligned InAs/InSb NWs

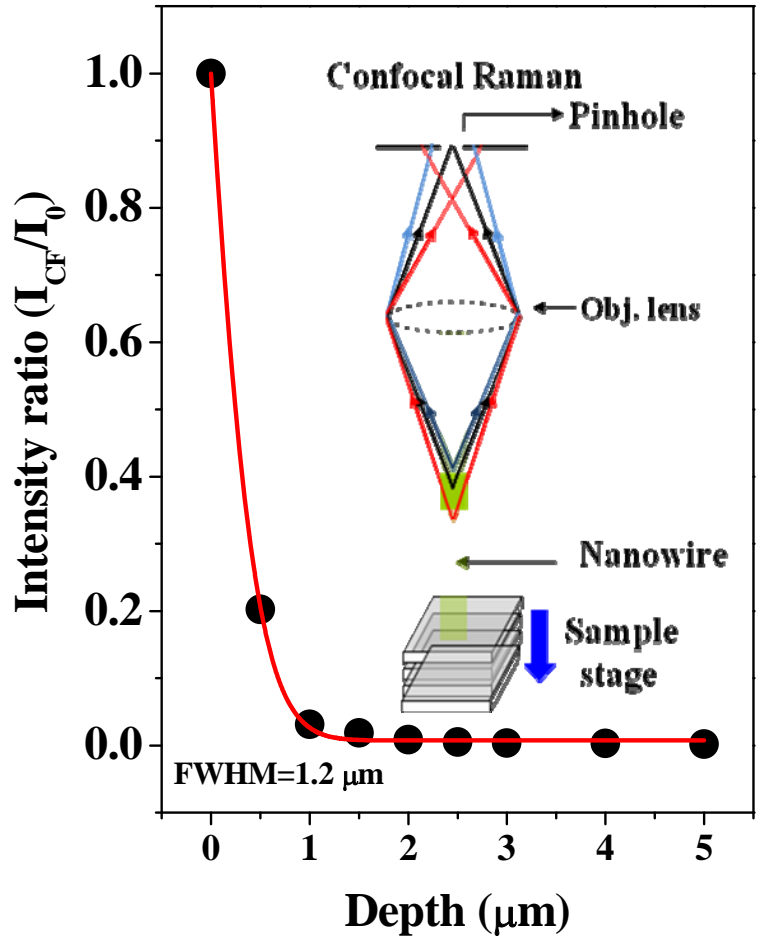

FIG. 2. The ratio of intensity of Raman scattered light through the pinhole ($I_{CF}$) and without the pinhole ($I_0$) as a function of depth from the focal plane. Inset of the fig is the cartoon diagram explaining confocal measurements on NW (shown by the green bar). Only the black rays, from dark segment, pass the pinhole and reach the detector. The blue and red rays, from lighter segments of the NW, are blocked by the pinhole.

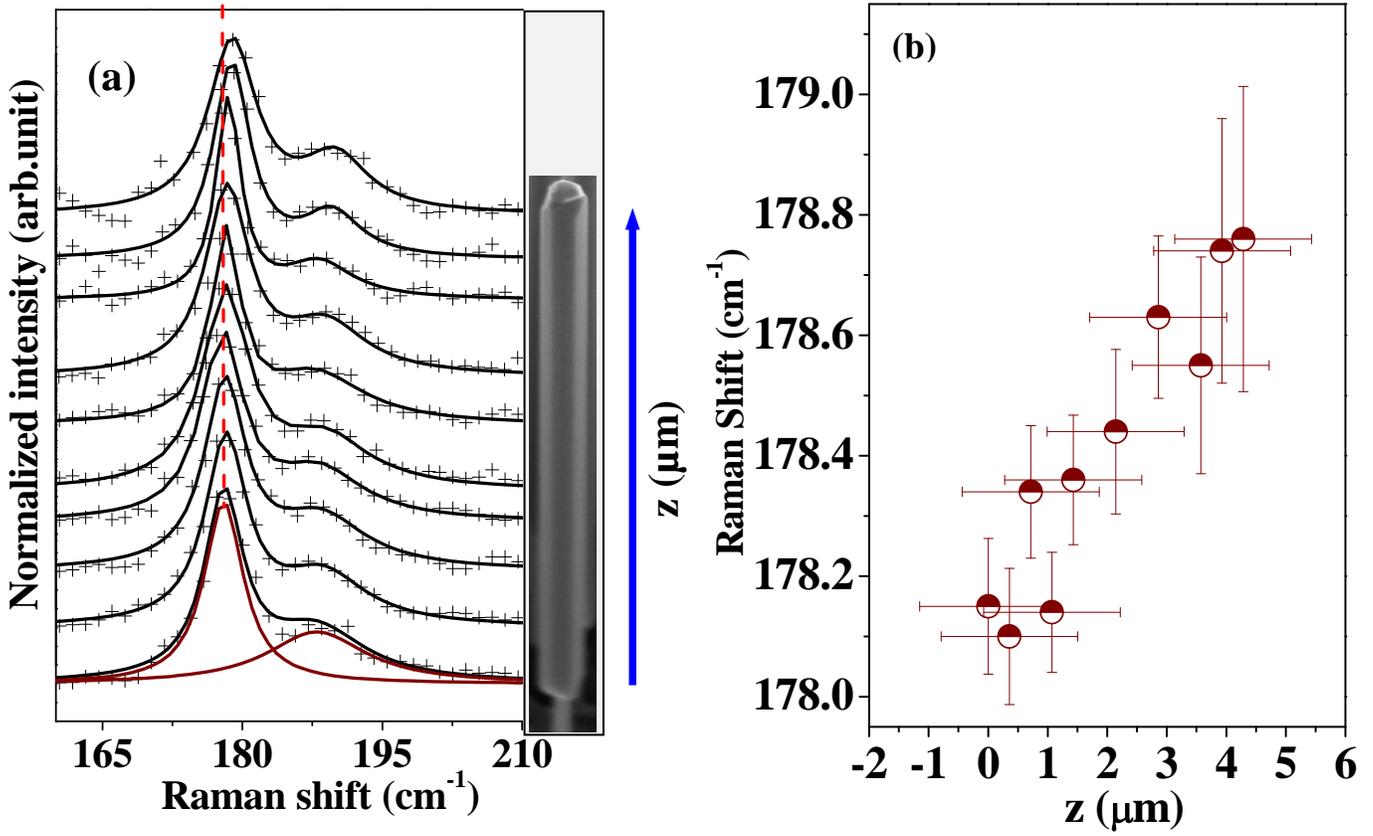

FIG. 3. (a) The evolution of Raman spectra along the axis of InAs/InSb NWs. The arrow marks the spectra, recorded from the bottom towards the tip of the NW. The vertical dashed line is drawn to show the shift in TO phonon frequency along the length of the NW. (b) The variation of the InSb TO phonon wavenumber along the length of the NW and horizontal error bar in each point indicates the step resolution.

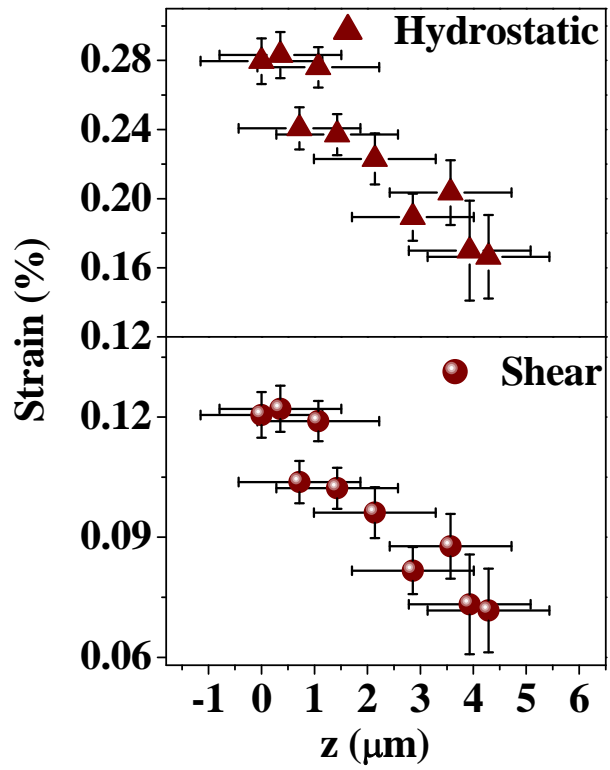

Fig. 4. Estimated variation of hydrostatic and shear strain along the axis of InAs/InSb NWs.

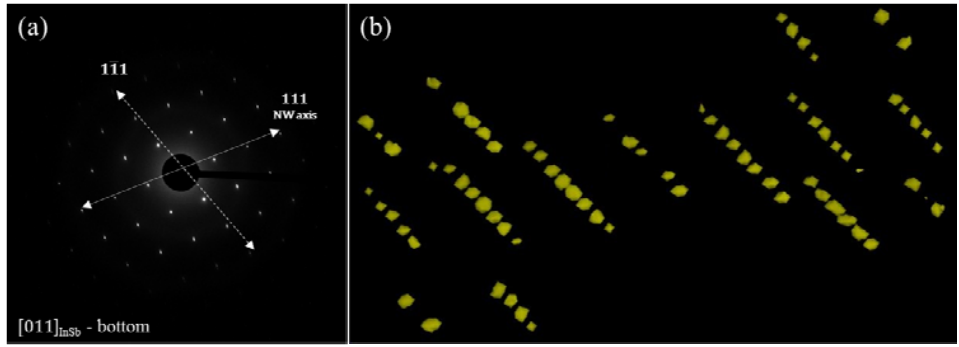

Fig. 5. (a) SAED pattern taken at the bottom of an InSb segment. Arrows show the measured distances between 333 and -3-3-3 spots used for estimating $d_{111}$ along the two equivalent directions. (b) 3-dimensional reconstruction of the reciprocal space of the InAs/InSb NW.